\documentclass[12pt]{article}

\usepackage{scicite}

\usepackage{times}

\usepackage{amsmath,amssymb}
\usepackage{siunitx}
\usepackage{graphicx}
\usepackage[version=3]{mhchem}

\newenvironment{sciabstract}{%
\begin{quote} \bf}
{\end{quote}}

\title{Optical trapping of nanoparticles in superfluid helium}

\author
{Yosuke Minowa,$^{1,2\ast}$ Xi Geng,$^{1}$ Keisuke Kokado,$^{1}$ Kentaro Sato,$^{3}$\\
Tatsuya Kameyama$^{2,3}$ Tsukasa Torimoto$^{3}$ Masaaki Ashida$^{1}$\\
\\
\normalsize{$^{1}$Graduate School of Engineering Science, Osaka University,}\\
\normalsize{1-3, Machikane-yama, Toyonaka, Osaka, Japan}\\
\normalsize{$^{2}$JST, PRESTO, 4-1-8 Honcho, Kawaguchi, Saitama, Japan}\\
\normalsize{$^{3}$Graduate School of Engineering, Nagoya University, Furo-cho, Chikusa-ku, Nagoya, Japan}\\
\\
\normalsize{$^\ast$To whom correspondence should be addressed; E-mail:  minowa@mp.es.osaka-u.ac.jp.}
}

\date{}

\begin{document} 


\baselineskip24pt


\maketitle


\begin{sciabstract}
Optical tweezers, the three-dimensional confinement of a nanoparticle by a strongly focused beam of light, have been widely employed in investigating biomaterial nanomechanics, nanoscopic fluid properties, and ultrasensitive detections in various environments such as inside living cells, at gigapascal pressure, and under high vacuum.  However, the cryogenic operation of solid-state-particle optical tweezers is poorly understood. In this study, we demonstrate the optical trapping of metallic and dielectric nanoparticles in superfluid helium below 2 K, which is two orders of magnitude lower than in the previous experiments. We prepare the nanoparticles via in-situ laser ablation. The nanoparticles are stably trapped with a single laser beam tightly focused in the superfluid helium. Our method provides a new approach for studying nanoscopic quantum hydrodynamic effects and interactions between quantum fluids and classical objects.
\end{sciabstract}

\textbf{Teaser} Optical tweezers can grab nanoparticles at extremely low temperature (1.4 K), showcasing cryogenic optical tweezers.


\section*{Introduction}

Superfluid helium is a peculiar quantum fluid appearing at low temperature (below \SI{2.17}{K}), and it comprises a non-viscous superfluid component with a viscous normal fluid component. The quantum nature of superfluid helium manifests itself in macroscopic phenomena such as unusually efficient heat transport, film flow, and vortex quantization. Owing to the higher transition temperature, a significantly larger number of superfluid helium atoms ($N\sim 10^{25}$) can be prepared by just pumping helium vapor, which is advantageous compared with other superfluids, such as Bose-Einstein condensates of cold atoms($N\lesssim 10^{7}$) that requires the complicated experimental techniques such as laser cooling. The large heat capacity ensures that any nanomaterial injected into the superfluid helium is quickly cooled and thermalized. This enables exploring the unprecedented interactions between quantum fluids and classical nanomaterials and nanoscopic quantum hydrodynamic effects\cite{liBrownianMotionSuperfluid2011,balazsBrownianMotionMirror1958a,moroshkinDynamicsVortexParticleComplexes2019}. A major obstacle in such approaches is the lack of experimental techniques to suspend/levitate the target nanoobject in quantum fluid, and controlling and probing the object motion precisely.

In this study, we demonstrate the stable suspension of nanoparticles in superfluid helium using optical trapping. Generally, optical trapping requires high numerical objective (NA) lenses\cite{ashkinObservationSinglebeamGradient1986,ashkinOpticalTrappingManipulation1987a} comprising several optical components to realize suitable optical potentials. Irrespective of the research design, this is a desirable condition in biomaterial manipulation\cite{ashkinOpticalTrappingManipulation1987a,ashkinOpticalTrappingManipulation1987,svobodaDirectObservationKinesin1993}, micro hydrodynamic studies\cite{frieseOpticalAlignmentSpinning1998,liMeasurementInstantaneousVelocity2010}, and trapping in extreme conditions\cite{liMillikelvinCoolingOptically2011,millenNanoscaleTemperatureMeasurements2014,ricciAccurateMassMeasurement2019,bowmanOpticalTrappingGigapascal2013,ashkinOpticalLevitationHigh1976,gieselerSubkelvinParametricFeedback2012, ahnUltrasensitiveTorqueDetection2020,rahmanLaserRefrigerationAlignment2017a,aritaLaserinducedRotationCooling2013,aritaCoherentOscillationsLevitated2020}. However, cryogenic conditions prohibit the use of multi-element optics, due to the large thermal deformation and fragility of the bonding and housing materials. Optical-levitation-trap schemes have been proposed previously\cite{ mundOpticalLevitationSingle2003}, where a long-working-distance and low-NA lens is placed outside cryostats to achieve optical trapping in low-temperature environments ($\sim \SI{180}{\kelvin}$). In such techniques, an upward-directed laser beam pushes microparticles, thereby balancing the gravitational force. The low-NA results in the relatively weak transverse optical force, which makes it difficult to trap smaller nanoparticles. In this study, we use a high-NA moulded aspheric lens to implement simple all-optical single beam trapping\cite{ashkinObservationSinglebeamGradient1986}. The lens was placed in the liquid helium cryostat and immersed in superfluid helium as shown in Fig. \ref{fig:particles}A. The strong gradient optical force combined with the very small thermal fluctuation energy ($\sim$ \SI{e-4}{\eV}) realize stable optical trapping of solid nanoparticles in superfluid helium. Our study demonstrates the optical trapping at 1.4 K, which is two orders of magnitude lower than previously reported
\cite{mundOpticalLevitationSingle2003,ishizakaSituObservationsFreezing2011a}. The all-optical method enables us to probe the suspended particle properties, such as the size of the particles. Moreover, the method would also provide an unprecedented scheme to study the nanoscopic quantum hydrodynamic phenomena such as Brownian motion in the quantum fluid\cite{liBrownianMotionSuperfluid2011}.

\section*{Results}
\subsection*{Calculation of optical force and effective potential energy}
\begin{figure}[h]
\includegraphics[width=\columnwidth]{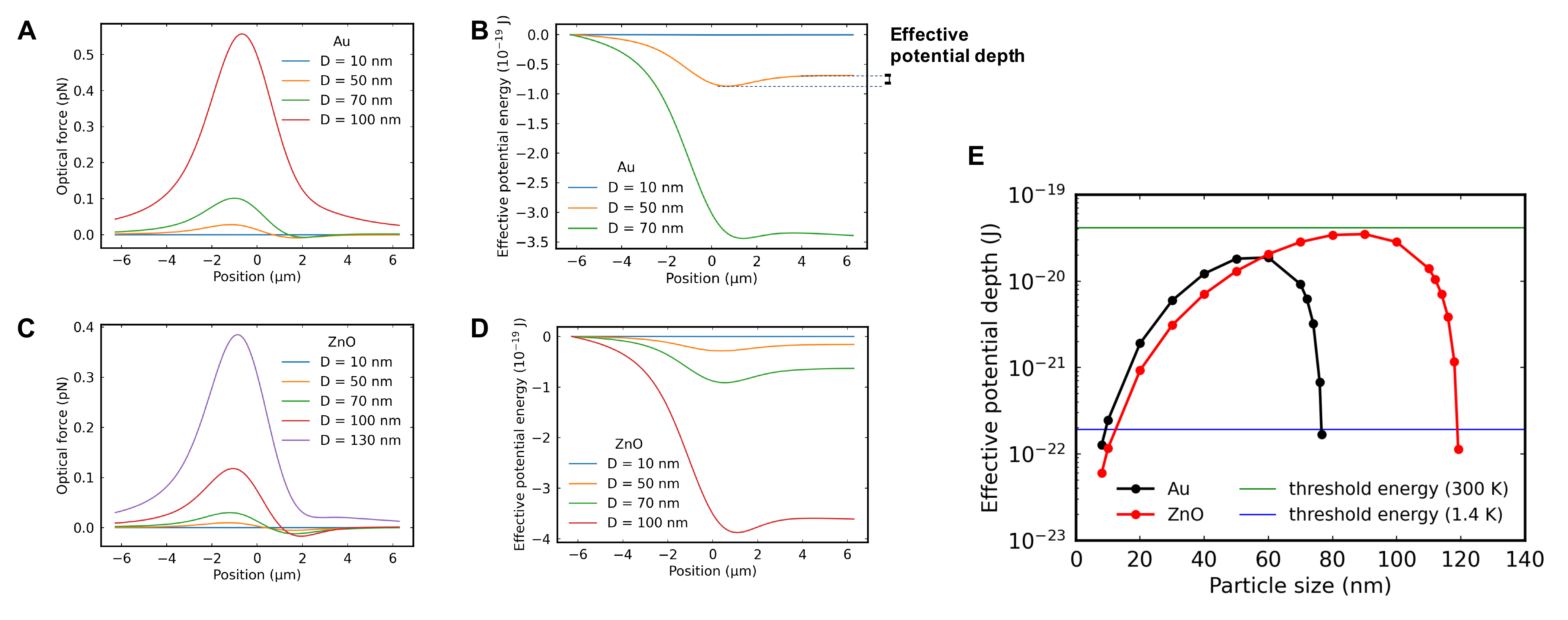} 
\caption{\textbf{Numerically calculated optical force and effective potential energy for the optical trapping threshold diameters estimation.} \textbf{(A, C)} Axial optical force versus the axial position of the particle with different diameters. \textbf{(B, D)} Corresponding effective potential energy curves. The data are shown only for the sizes in equilibrium. \textbf{(E)} Depth of the effective potential as a function of the particle size with the stability threshold line, which is 10 times the thermal energy (green line for $T=\SI{300}{\kelvin}$ and blue line for $T=\SI{1.4}{\kelvin}$).}
\label{fig:calculation}
\end{figure}

Figure \ref{fig:calculation}A shows the predicted axial optical force, which is along the x-axis, and the light propagation direction; the forces shown in the figure were calculated using the parameters matched to the experimental setup, enabling us to investigate whether the stable optical trapping in superfluid helium is possible. The x-component of the optical force exerted on a gold nanoparticle is shown as a function of the nanoparticle position x on the optical axis (see Fig. \ref{fig:particles}A for the coordinate system). Gold nanoparticles are widely used in optical trapping, owing to the large polarizability \cite{svobodaOpticalTrappingMetallic1994} that enables stable optical trapping. The optical force is calculated based on the generalized Lorenz–Mie theory\cite{nieminenOpticalTweezersComputational2007} for different particle sizes. The calculated optical force is 2 to 3 orders of magnitude higher than the gravitational force; therefore the gravitational effect is negligible here. Smaller particles remain at equilibrium, where the optical force is zero and the slope of the curve is negative. This ensures that the particles experience a force toward the equilibrium point. This can be seen more clearly in Fig \ref{fig:calculation}B, where we show the effective potential energy by integrating the force curve. Note that the optical force normally includes a non-conservative force\cite{divittCancellationNonconservativeScattering2015} and the calculated effective potential energy is just used for estimating the stability of the optical trapping. The existence of the potential minimum indicates the possibility of stable optical trapping for the smaller particles. Figure \ref{fig:calculation}A reveals that there is no equilibrium point for the larger particles. The overall size dependency is consistent with the Rayleigh scattering cross section of the nanoparticles, and thus, the ’pushing’ scattering force scales with the sixth power of the particle size\cite{jacksonClassicalElectrodynamics1998}. In contrast, the optical dipole force scales with the third power of the particle size\cite{haradaRadiationForcesDielectric1996}, which pulls the particle toward the focus of the light. Therefore, the scattering force is dominant for the larger particles, whereas for the smaller particles, the optical dipole force becomes dominant.

The particle size dependence of the effective potential energy depth is depicted in Fig. \ref{fig:calculation}E. It has been established that stable optical trapping requires the potential depth to be larger than $10\times k_B T$. The horizontal green line indicates the corresponding threshold energy for $T = \SI{300}{\kelvin}$, showing the difficulty of the aspheric-lens-based optical trapping near room temperature. However, much lower threshold energy for superfluid helium (blue line in Fig. \ref{fig:calculation}E) enables us the stable optical trapping. The lower threshold size of the particle is $\SI{10}{nm}$ and the upper threshold size of the particles is $\SI{77}{nm}$. We also explore the possibility of optical trapping in superfluid helium for dielectric nanoparticles. Zinc oxide is a high-refractive index material with larger polarizability. Figure \ref{fig:calculation}C and D respectively show the corresponding optical force and effective potential energy curves for zinc oxide. The resulting effective potential energy depth shown in Fig. \ref{fig:calculation}E indicates that stable optical trapping is possible also for zinc oxide nanoparticles of size $\SI{10}\sim \SI{120}{nm}$.

\subsection*{Particle loading}
\begin{figure}[h]
\includegraphics[width=1.0\columnwidth]{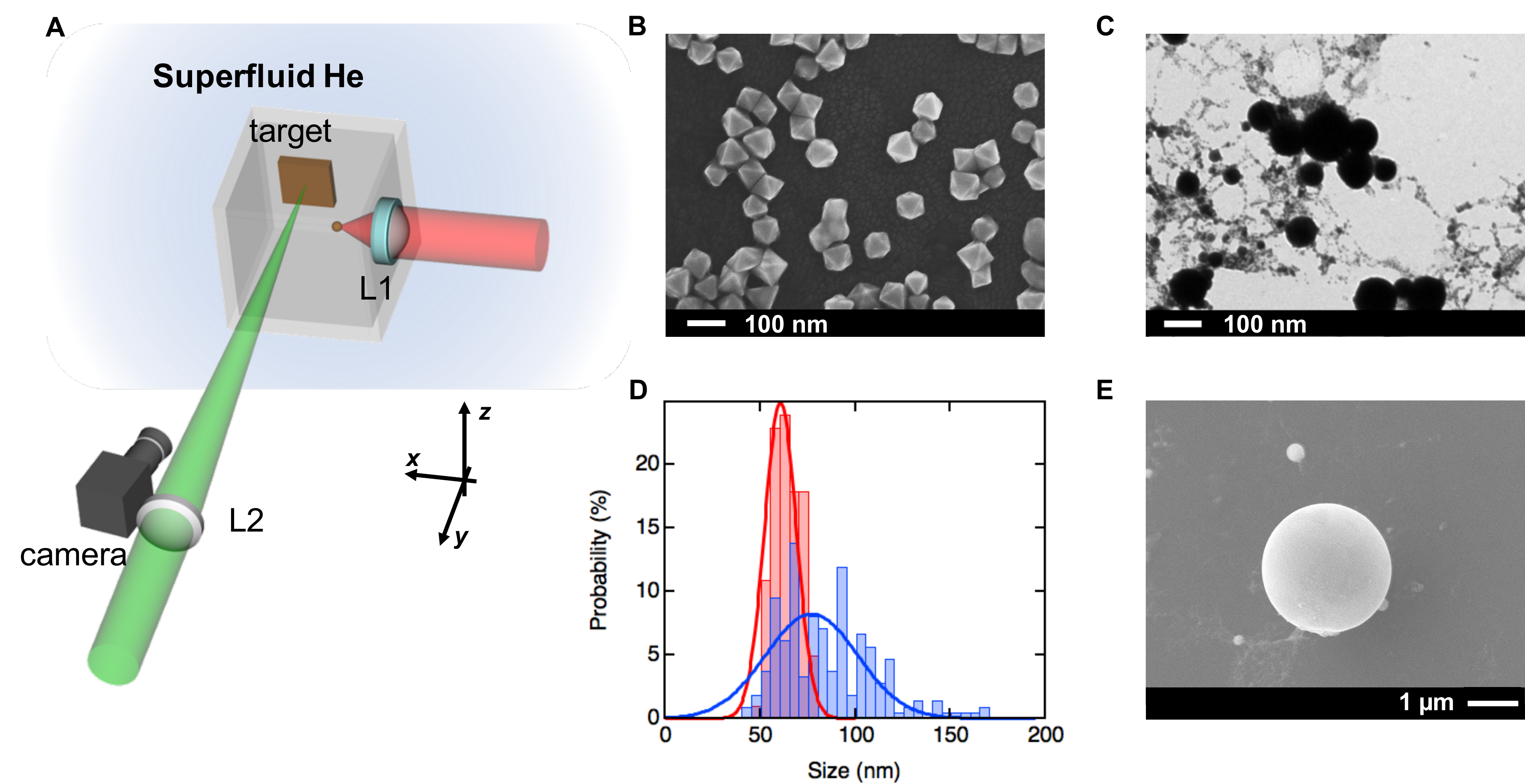}
\caption{\textbf{Optical trapping of gold and zinc oxide nanoparticles in superfluid helium.} \textbf{(A)} A linearly polarized laser beam is tightly focused with an aspheric lens (L1) immersed in superfluid helium. The target nanoparticles are loaded in superfluid helium via  laser ablation with a focusing lens (L2). \textbf{(B)} SEM image of octahedral gold nanoparticles before laser ablation. \textbf{(C)} TEM image of gold nanoparticles ejected due to laser ablation. \textbf{(D)} Gold nanoparticle size distribution before (red) and after (blue) laser ablation. \textbf{(E)} Zinc oxide micro and nanoparticles synthesized by laser ablation.}
\label{fig:particles}
\end{figure}

A major issue in nanoparticle optical trapping is preparing slow nanoparticles near the focal point. In particular, cryogenic conditions prohibit the use of the standard procedures such as the "go and pick" scheme\cite{ashkinOpticalTrappingManipulation1987a} or using aerosols \cite{minowaOpticalLevitationMicrodroplet2015} to prepare nanoparticles. 
In this study, we utilize the laser ablation technique to load the nanoparticles. Gold nanoparticles (Methods) are coated on a glass coverslip, mounted in the liquid helium cryostat, and immersed in superfluid helium. The coverslip is irradiated by a pulsed laser light, which releases gold nanoparticles. Figure \ref{fig:particles}B and C are typical scanning and transmission electron microscope (SEM and TEM) images before and after laser ablation-induced ejection. Noticeable shape changes from octahedrons to spheres indicate that the ejection process accompanies melting and/or vaporization. 

Figure \ref{fig:particles}D shows the particle size distribution before and after laser ablation. The size distributions before (after) the laser ablation are well-fitted using a Gaussian function with a mean of $\SI{60}{nm}$ ($\SI{77}{nm}$) and standard deviation of $\SI{8}{nm}$ ($\SI{24}{nm}$). Although the size distribution of the ejected gold nanoparticles is broader than that of the original gold nanoparticles, half of the ejected particles lie within the range of the stability-size regime $\SI{10}\sim \SI{77}{nm}$ illustrated in Fig. \ref{fig:calculation}E.

Dielectric nanoparticles can be synthesized in situ using laser ablation in the superfluid helium. Figure \ref{fig:particles}E shows the typical SEM image of the synthesized zinc oxide particles. The zinc oxide particles are highly spherical \cite{minowaInnerStructureZnO2017}, which is suitable for optical trapping. Many synthesized zinc oxide particles have dimensionalities in the stability-size regime, although its difficult to control the synthesized particle size distribution. This is in contrast with the case of gold nanoparticles, where the loaded-particle size distribution reflects the original size distribution.

Figure \ref{fig:result}A illustrates the experimental arrangement. A gold-nanoparticle-coated coverslip is placed in a cuvette in the liquid helium cryostat. A lens for optical trapping (L1) is mounted in the middle of the cuvette sidewall. The cuvette has small gaps between the walls, ensuring the flow of liquid helium to and from the cuvette. After the liquid helium is transferred to the cryostat, the fluid is cooled using a vacuum pump, and the temperature falls below the superfluid transition temperature. We maintain this temperature $\SI{1.4}{\kelvin}$ during the experiment. The irradiation of a nanosecond laser pulse on the coverslip initiates the gold nanoparticle ejection process, and the ejected nanoparticles are dispersed in the cuvette. The gold nanoparticles exhibit random motion until landing on the cuvette surface or exiting the cuvette. 

\subsection*{Optical trapping in superfluid helium}
\begin{figure}[h]
\includegraphics[width=1.0\columnwidth]{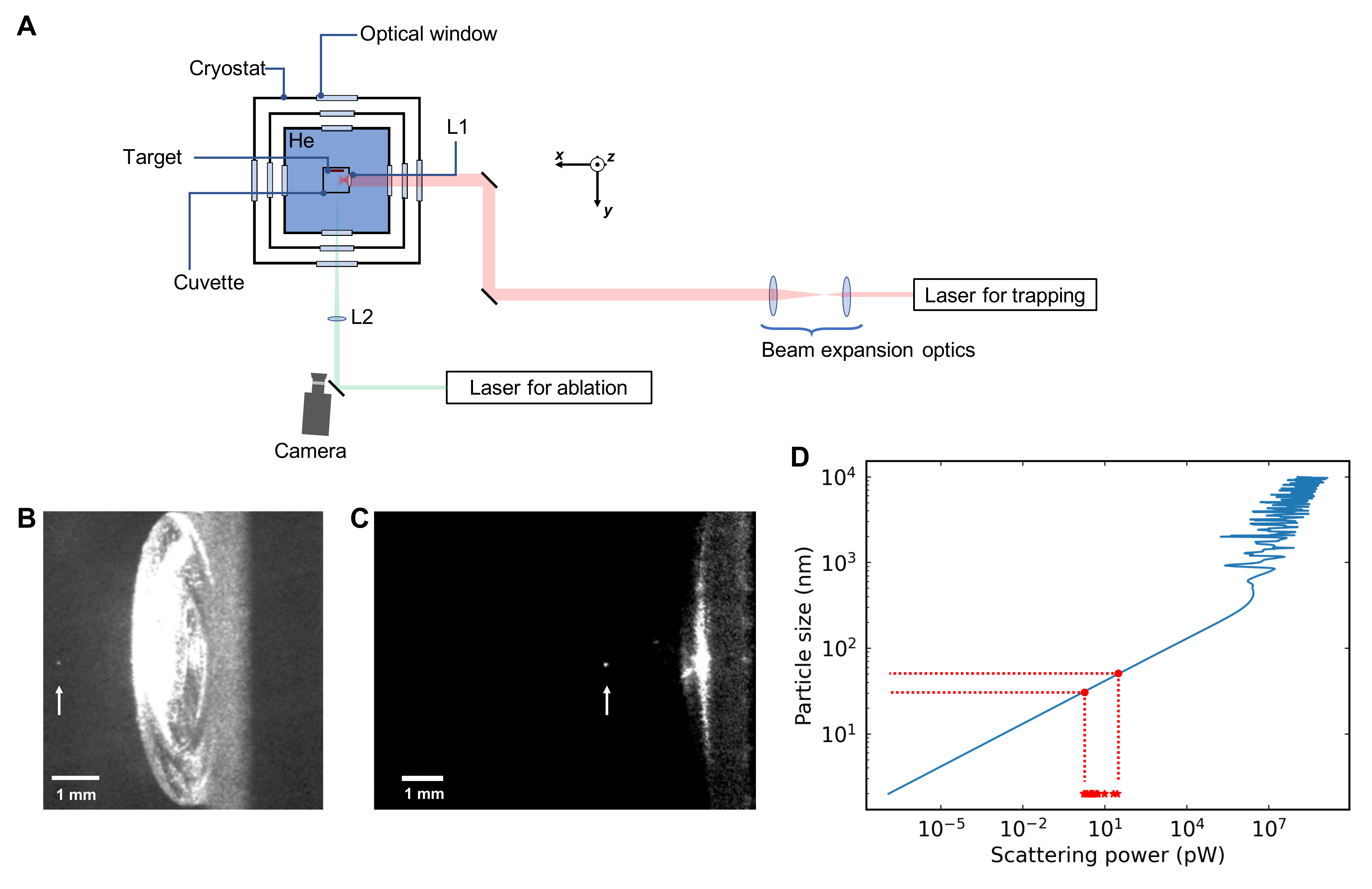}
\caption{\textbf{Optical trapping of nanoparticles in superfluid helium.} \textbf{(A)} Schematic of the experiment. A pulsed nanosecond-laser beam is focused with a lens (L2) onto a target substrate. The loaded/synthesized nanoparticles are dispersed in the superfluid helium and are optically trapped by a linearly polarized beam of light focused with a lens (L1) immersed in the superfluid helium. Light scattering from the optically trapped (\textbf{B}) gold and (\textbf{C}) zinc oxide nanoparticle is imaged through the optical windows of the cryostat, and recorded by a CMOS camera. \textbf{(D)} Estimated trapped particle size as a function of the scattering light power. The blue curve corresponds to the calculated relation between the particle size and the scattering power. The red star marks indicate the experimentally detected scattering powers. The red dotted lines correspond to the eye-guides.}
\label{fig:result}
\end{figure}

Once a nanoparticle reaches the focal point of the tightly focused beam of light, it gets optically trapped. The strong light scattering enables visualizing the trapped nanoparticles, as shown in Fig. \ref{fig:result}B. In addition, the zinc oxide nanoparticles can be optically trapped by replacing the gold-nanoparticle-coated coverslip with the zinc oxide bulk target. In situ laser-ablation synthesis allows dispersing zinc oxide nanoparticles in superfluid helium. The nanoparticles are captured by the optical force and stably trapped thereafter. The trapped nanoparticles are very stable and can remain in the trap typically for over 30 min. If we tentatively block the laser beam, the trapped particle escapes from the trapping site, ensuring that the trapping is truly due to the optical force. An animated visualization is provided as Supplementary Movie.

Our method provides an optical-tweezers-based-approach\cite{liMeasurementInstantaneousVelocity2010, frieseOpticalAlignmentSpinning1998} to study nanoscopic quantum hydrodynamics and interaction between quantum fluids and classical nanoobjects\cite{liBrownianMotionSuperfluid2011,balazsBrownianMotionMirror1958a,moroshkinDynamicsVortexParticleComplexes2019}. Determining the trapped particle size is very necessary because the size governs the interaction between the particle and the surrounding quantum fluid. In our study, zinc oxide particles of sizes ranging from nanometres to micrometres\cite{minowaInnerStructureZnO2017} were fabricated through the laser ablation process. We demonstrate the size determination of the actual trapped particle using the Mie scattering theory\cite{bohrenAbsorptionScatteringLight1983}. The blue curve in Fig. \ref{fig:result}D corresponds to the theoretically calculated relation between the particle size and the optical power scattered from the trapped particle into the observation solid angle. The red star marks in the figure indicate the detected actual scattering powers obtained from separate trapping events (see Supplementary Information). Note that the values are corrected in terms of the total detection efficiency, including the transmissivity of the detection optics. The scattering powers are well-below the Mie resonance region starting from $\sim \SI{1e5}{pW}$, where the relation between the particle size and the scattering power is not straightforward. In other words, in our experimental condition, Rayleigh scattering is considered a good approximation, where the scattering power is proportional to the sixth power of the particle size. Accordingly, the curve is almost straight in the log–log plot. The trapped particle size was determined to be $30 \sim 50\, \si{nm}$ by comparing the experimental data and the theoretical curve, which exactly matches the calculated stability size range.

\section*{Discussion}
The techniques developed in this study enable trapping and manipulating nanoparticles in superfluid helium and also characterizing the trapped particle all-optically. Our method will open up a new pathway to study quantum hydrodynamics, as our optical trapping scheme can be extended to monitor trapped nanoparticle positions with high spatial and temporal resolution\cite{liMeasurementInstantaneousVelocity2010}, enabling the observation of the Brownian motion in the quantum fluid\cite{liBrownianMotionSuperfluid2011}.  Another possibility is combining our method with existing ingenious techniques for superfluid helium research. One exciting example is the quantum vortex visualization in superfluid helium\cite{bewleySUPERFLUIDHELIUMVisualization2006}. When a bunch of tracer nanoparticles is dispersed in superfluid helium, the particles are stabilized along the quantum vortex line. This is because the quantum vortex core is a local minimum of the pressure field. Therefore, quantum vortex dynamics can be visualized by imaging the scattered light from the nanoparticles, leading to recent experiments and studies on the hidden nature of the vortex-vortex interaction\cite{bewleyCharacterizationReconnectingVortices2008}. These remarkable experiments, however, rely on the observation of the free motion of the quantum vortex, and the accidental collision of two vortices. Our results exhibit a striking possibility: optical trapping and control of the composite quantum vortex and tracer particle system. This technique could act as a new method to control the quantum vortex motion, and dynamically perturb and excite the quantum vortex states.

\section*{methods}

\subsection*{Synthesis of octahedral gold nanoparticles}
Octahedral gold nanoparticles were synthesized in solution via a previously reported method with a modification\cite{liFacilePolyolRoute2008}. A 20 wt\% poly (diallyldimethylammonium chloride) aqueous solution (Mw = 100000-200000) (\SI{80}{mm^3}) and a \SI{0.1}{mol.dm^{-3}} HCl aqueous solution (\SI{80}{mm^3}) were injected into diethylene glycol (\SI{4.0}{cm^3}) in a test tube with magnetic stirring at \SI{30}{\degreeCelsius}. A \SI{19}{mm^3} portion of \SI{0.11}{mol.dm^{-3}} \ce{HAuCl4} aqueous solution was added into the mixture, followed by heat treatment at \SI{230}{\degreeCelsius} for 60 min with vigorous stirring in a \ce{N2} atmosphere. After cooling to room temperature, the synthesized octahedral gold nanoparticles were separated from the solution by adding acetone (\SI{4.0}{cm^3}), followed by centrifugation at 15000 rpm for 10 min. The precipitates of gold nanoparticles were dispersed in water, followed by centrifugation (15000 rpm, 10 min). This washing procedure was repeated three times. Finally, the purified octahedral gold nanoparticles were dispersed again in \SI{1.0}{cm^3} water.

\subsection*{Optical trapping in cryostat}
We placed a \SI{3}{cm} x \SI{3}{cm} x \SI{3}{cm} cuvette in superfluid helium. The cuvette sidewall has a mounting hole for the aspheric lens for optical trapping. The experimental process occurred entirely in this cuvette. The cuvette has small gaps between the walls, ensuring the flow of liquid helium to and from the cuvette. We mounted a coverslip coated with the gold nanoparticles and a bulk sintered semiconductor zinc oxide target in the cuvette. The liquid helium temperature is maintained at approximately 1.4 K during the experiment. Nanosecond light pulses from a frequency-doubled Q-switched Nd:YAG laser (wavelength 532 nm, pulse duration 10 ns, repetition rate 10 Hz, and pulse energy 1 mJ) were focused onto the target surface with a spot size of $\sim$\SI{40}{\micro \meter}, using a lens of 200 mm focal length. The ejected particles were optically trapped using continuous-wave laser (wavelength 785 nm). The typical power of the laser for the trapping is $\sim \SI{100}{mW}$, ranging from $\SI{50}{mW}$ to $\SI{500}{mW}$.



\bibliography{hetrap2}

\begin{thebibliography}{10}

\bibitem{liBrownianMotionSuperfluid2011}
X.~Li, R.~Cheng, T.~Li, Q.~Niu, {\it arXiv:1107.0485\/}  (2011).

\bibitem{balazsBrownianMotionMirror1958a}
N.~L. Balazs, {\it Physical Review\/} {\bf 109}, 232 (1958).

\bibitem{moroshkinDynamicsVortexParticleComplexes2019}
P.~Moroshkin, P.~Leiderer, K.~Kono, S.~Inui, M.~Tsubota, {\it Physical Review
  Letters\/} {\bf 122}, 174502 (2019).

\bibitem{ashkinObservationSinglebeamGradient1986}
A.~Ashkin, J.~M. Dziedzic, J.~E. Bjorkholm, S.~Chu, {\it Optics Letters\/} {\bf
  11}, 288 (5 01, 1986).

\bibitem{ashkinOpticalTrappingManipulation1987a}
A.~Ashkin, J.~M. Dziedzic, T.~Yamane, {\it Nature\/} {\bf 330}, 769 (1987).

\bibitem{ashkinOpticalTrappingManipulation1987}
A.~Ashkin, J.~M. Dziedzic, {\it Science\/} {\bf 235}, 1517 (1987).

\bibitem{svobodaDirectObservationKinesin1993}
K.~Svoboda, C.~F. Schmidt, B.~J. Schnapp, S.~M. Block, {\it Nature\/} {\bf
  365}, 721 (1993).

\bibitem{frieseOpticalAlignmentSpinning1998}
M.~E.~J. Friese, T.~A. Nieminen, N.~R. Heckenberg, H.~{Rubinsztein-Dunlop},
  {\it Nature\/} {\bf 394}, 348 (7 23, 1998).

\bibitem{liMeasurementInstantaneousVelocity2010}
T.~Li, S.~Kheifets, D.~Medellin, M.~G. Raizen, {\it Science\/} {\bf 328}, 1673
  (2010).

\bibitem{liMillikelvinCoolingOptically2011}
T.~Li, S.~Kheifets, M.~G. Raizen, {\it Nat Phys\/} {\bf 7}, 527 (online 3 20,
  2011).

\bibitem{millenNanoscaleTemperatureMeasurements2014}
J.~Millen, T.~Deesuwan, P.~Barker, J.~Anders, {\it Nature Nanotechnology\/}
  {\bf 9}, 425 (2014).

\bibitem{ricciAccurateMassMeasurement2019}
F.~Ricci, M.~T. Cuairan, G.~P. Conangla, A.~W. Schell, R.~Quidant, {\it Nano
  Letters\/}  (2019).

\bibitem{bowmanOpticalTrappingGigapascal2013}
R.~W. Bowman, G.~M. Gibson, M.~J. Padgett, F.~Saglimbeni, R.~Di~Leonardo, {\it
  Physical Review Letters\/} {\bf 110}, 095902 (2 28, 2013).

\bibitem{ashkinOpticalLevitationHigh1976}
A.~Ashkin, J.~M. Dziedzic, {\it Applied Physics Letters\/} {\bf 28}, 333
  (1976).

\bibitem{gieselerSubkelvinParametricFeedback2012}
J.~Gieseler, B.~Deutsch, R.~Quidant, L.~Novotny, {\it Physical Review
  Letters\/} {\bf 109}, 103603 (9 7, 2012).

\bibitem{ahnUltrasensitiveTorqueDetection2020}
J.~Ahn, {\it et~al.\/}, {\it Nature Nanotechnology\/} {\bf 15}, 89 (2020).

\bibitem{rahmanLaserRefrigerationAlignment2017a}
A.~T. M.~A. Rahman, P.~F. Barker, {\it Nature Photonics\/} {\bf 11}, 634
  (2017).

\bibitem{aritaLaserinducedRotationCooling2013}
Y.~Arita, M.~Mazilu, K.~Dholakia, {\it Nature Communications\/} {\bf 4}, 2374
  (8 28, 2013).

\bibitem{aritaCoherentOscillationsLevitated2020}
Y.~Arita, S.~H. Simpson, P.~Zem{\'a}nek, K.~Dholakia, {\it Science Advances\/}
  {\bf 6}, eaaz9858 (2020).

\bibitem{mundOpticalLevitationSingle2003}
C.~Mund, R.~Zellner, {\it ChemPhysChem\/} {\bf 4}, 630 (2003).

\bibitem{ishizakaSituObservationsFreezing2011a}
S.~Ishizaka, T.~Wada, N.~Kitamura, {\it Chemical Physics Letters\/} {\bf 506},
  117 (2011).

\bibitem{svobodaOpticalTrappingMetallic1994}
K.~Svoboda, S.~M. Block, {\it Optics Letters\/} {\bf 19}, 930 (7 01, 1994).

\bibitem{nieminenOpticalTweezersComputational2007}
T.~A. Nieminen, {\it et~al.\/}, {\it Journal of Optics A: Pure and Applied
  Optics\/} {\bf 9}, S196 (2007).

\bibitem{divittCancellationNonconservativeScattering2015}
S.~Divitt, L.~Rondin, L.~Novotny, {\it Optics Letters\/} {\bf 40}, 1900 (2015).

\bibitem{jacksonClassicalElectrodynamics1998}
J.~D. Jackson, {\it Classical {{Electrodynamics}}\/} ({Wiley}, 1998), third
  edn.

\bibitem{haradaRadiationForcesDielectric1996}
Y.~Harada, T.~Asakura, {\it Optics Communications\/} {\bf 124}, 529 (3 15,
  1996).

\bibitem{minowaOpticalLevitationMicrodroplet2015}
Y.~Minowa, R.~Kawai, M.~Ashida, {\it Optics Letters\/} {\bf 40}, 906 (2015).

\bibitem{minowaInnerStructureZnO2017}
Y.~Minowa, Y.~Oguni, M.~Ashida, {\it Optics Express\/} {\bf 25}, 10449 (2017).

\bibitem{bohrenAbsorptionScatteringLight1983}
C.~F. Bohren, D.~R. Huffman, {\it Absorption and Scattering of Light by Small
  Particles\/} ({Wiley}, 1983).

\bibitem{bewleySUPERFLUIDHELIUMVisualization2006}
G.~P. Bewley, D.~P. Lathrop, K.~R. Sreenivasan, {\it Nature\/} {\bf 441}, 588
  (2006).

\bibitem{bewleyCharacterizationReconnectingVortices2008}
G.~P. Bewley, M.~S. Paoletti, K.~R. Sreenivasan, D.~P. Lathrop, {\it
  Proceedings of the National Academy of Sciences\/} {\bf 105}, 13707 (2008).

\bibitem{liFacilePolyolRoute2008}
C.~Li, K.~L. Shuford, M.~Chen, E.~J. Lee, S.~O. Cho, {\it ACS Nano\/} {\bf 2},
  1760 (2008).

\end{thebibliography}

\bibliographystyle{Science}

\section*{Acknowledgments}
\textbf{Funding:} This work was supported by the MEXT/JSPS KAKENHI Grant Number JP17K17841, JP16H06505, JP16H06507, and JP18KK0387 and by JST, PRESTO Grant Number JPMJPR18T5 and JPMJPR1909, Japan.
\textbf{Author contributions:} Y.M. conceived and designed the project and wrote the paper. Y.M., X.G and K.K. performed the experiments and analyzed the data. K.S., T.K., and T.T. synthesized gold nanoparticles and characterized them. M.A. provided technical support, and assisted in writing and editing the manuscript. \textbf{Competing interests:} The authors declare that they have no competing interests." If this is not accurate, please list the competing interests. \textbf{Data and materials availability:}  All data are available in the manuscript or supplementary materials.

\section*{Supplementary materials}
Supplementary Text\\
Movie. S1\\


\clearpage

\end{document}